\documentclass[twocolumn]{aastex63}
\usepackage{url}
\usepackage[utf8]{inputenc}
\usepackage{graphicx}
\usepackage{xcolor}

\newcommand\etal{~et~al.}

\usepackage[sort&compress]{natbib}

\graphicspath{{./}{figures/}}

\shorttitle{Lightcurve Evolution of the nearest TDE}
\shortauthors{Perlman et al.}

\begin{document}

\title{Lightcurve Evolution of the nearest Tidal Disruption Event:  A late-time, radio-only flare}

\author{Eric S. Perlman}
\affiliation{Department of Physics and Space Sciences, Florida Institute of Technology, 150 W. University Blvd., Melbourne, FL 32901, USA}

\author{Eileen T. Meyer}
\affiliation{Department of Physics, University of Maryland -- Baltimore County, 1000 Hilltop Circle, Baltimore, MD  21250, USA}

\author{Q. Daniel Wang}
\affiliation{Department of Astronomy, University of Massachusetts, LGRT-B 619E, 710 North Pleasant Street, Amherst, MA 01003-9305, USA} 

\author{Qiang Yuan}

\affiliation{Key Laboratory of Dark Matter and Space Astronomy, Purple Mountain Observatory, Chinese Academy of Sciences, 10 Yuanhua Road, Nanjing 210023, China}
\affiliation{School of Astronomy and Space Science, University of Science and Technology of China, Hefei 230026, China}

\author{Richard Henriksen, Judith Irwin}
\affiliation{Department of Physics, Engineering Physics \& Astronomy, Queens University, Kingston, Ontario, K7L 3N6, Canada}



\author{Jiangtao Li}
\affiliation{Department of Astronomy, University of Michigan, 311 West Hall, 1085 South University Avenue,  Ann Arbor, Michigan, 48109, USA}

\author{Theresa Wiegert}
\affiliation{Department of Physics, Engineering Physics \& Astronomy, Queens University, Kingston, Ontario, K7L 3N6, Canada}

\author{Haochuan Li}
\affiliation{Department of Physics, Engineering Physics \& Astronomy, Queens University, Kingston, Ontario, K7L 3N6, Canada}

\author{Yang Yang}
\affiliation{School of Astronomy and Space Science, Nanjing University, Nanjing 210093, China}

\begin{abstract}

Tidal disruption events (TDEs) occur when a star passes close enough to a 
galaxy's supermassive black hole to be disrupted by tidal forces. 
We discuss new observations of IGRJ12580+0134, a TDE observed in NGC 4845 
($d=17$~Mpc) in November 2010, with the  Karl G. Jansky {\sl Very Large Array} 
(VLA{\footnote{The National Radio Astronomy Observatory is a facility of the 
National Science Foundation operated under cooperative agreement by 
Associated Universities, Inc.}}).
We also discuss a reanalysis of 2010-2011 {\sl Swift} and {\sl XMM-Newton} observations, as well as 
new, late-time  {\sl Swift} observations. 
Our JVLA observations show a decay of the nuclear radio flux until 2015, when a plateau was seen, and then a significant ($\sim$factor 3) radio flare during 2016.  The 2016 radio flare was also accompanied by radio spectral changes, but was not seen in the X-rays. We model the flare as resulting from the interaction of the nuclear jet with a cloud in the interstellar medium.  This is distinct from late-time X-ray flares in a few other TDEs where changes in the accretion state and/or a fallback event were suggested, neither of which appears possible in this case. 
Our reanalysis of the {\sl Swift} and {\sl XMM-Newton} data from 2011  shows significant evidence for thermal emission from a disk, as well as a very soft power-law. This, in addition to the extreme X-ray flux increase seen in 2010 (a factor of $>$100) bolsters the identification of IGRJ12580+0134 as a TDE, not an unusual AGN variability event.

\end{abstract}



\section{Introduction}

Tidal disruption events (TDEs) are expected to occur every $10^3-10^5$
years for a typical galaxy \citep{Magorrian99,Wang04,vanVelzen14,Holoien16}. They occur when a star 
or sub-stellar object passes close enough by the galaxy's central supermassive 
black hole (SMBH) to be tidally disrupted. The debris of the disrupted
object gets accreted onto the black hole, producing flaring emission
at X-ray, ultraviolet, and optical wavelengths. Jets can also be launched
by the SMBH after a TDE. When they interact with the circum-nuclear medium 
(CNM) high energy particle acceleration could occur. Sw J1644+57 \citep[$z=0.3534$;][]{Bloom11,Burrows11,Levan11,Zauderer11,Zauderer13} 
and Sw J2058+05 \citep[$z=1.1853$;][]{Cenko12} are examples of TDEs 
that have associated jets, exhibiting super-Eddington X-ray emission and 
a long lasting radio emission expected to arise from the jet-CNM interaction. 
Detailed modeling of both Sw J1644+57 and Sw J2058+05 suggests that the jets 
were strongly relativistically beamed. A recent review by \citet{Alexander20} 
discusses the history of radio-loud TDE to date. It is natural to expect 
that there should be more events with off-axis jets, even though detection 
of these would be less likely, with statistics similar to the detection of 
off-axis blazars in a high flux limit survey \citep[e.g.,][]{Urry84}.

IGR J12580+0134 is a TDE that occurred in 2010 in the nucleus of NGC 4845 -- a galaxy 
located in the Virgo cluster.  Its distance of just $\sim17$ Mpc, gave us a rare 
chance to scrutinize a TDE and its aftermath with the highest possible resolution.
The source was initially detected in 2010 November by {\it Integral} \citep{Walter11}. 
Follow-up X-ray observations with {\sl XMM-Newton}, {\sl Swift} and {\sl MAXI}, 
together with {\it Integral} data, suggest that the source probably resulted from 
a tidal disruption of a super-Jupiter by the galaxy's central SMBH \citep[i.e., 
a sub-stellar TDE;][]{Nikolajuk13}. The source is underluminous compared to many 
TDEs (sub-Eddington as compared to super-Eddington).
The radio counterpart of the TDE was detected serendipitously in 2011 December 
by the {\sl VLA} in a nearby galaxy survey \citep[CHANG-ES;][]{Irwin15} where 
the core was a factor of more than 10 brighter than seen in {\it FIRST} observations 
conducted between 1993-2004. A search of archival {\it Planck} data showed 
bright flaring emissions in January 2011 in the millimeter wavelengths \citep{Yuan2016}.

The radio spectrum, peaking at GHz frequencies, and its evolution through 2015 
suggest self-absorbed synchrotron emission with changing optical thickness. 
{\sl VLBA} and {\sl VLA} observations of IGRJ1258+0134 in 2015 detected both
a bright core on arcsecond scales and resolved, milliarcsecond-scale 
emission \citep{Perlman17}. 
The resolved milli-arcsecond core emission was believed to be due to 
the expansion of the radio jet. A model for the radio jet \citep{Irwin15,Perlman17} 
shows the data to be consistent with an initial Lorentz factor $\Gamma_i\sim10$ 
and a viewing angle of $40^{\circ}$ \citep{Lei16}. 

Here we discuss new observations of IGRJ1258+0134 obtained with the {\sl VLA}, 
the Neil Gehrels {\sl Swift} X-ray Telescope and {\sl XMM-Newton}. The 2015-2016 
VLA data (Table 1) shows a flare in the radio that is {\it not} accompanied by a 
similar increase in the X-ray light curve.  A reanalysis of archival data from 
{\sl Swift} and {\sl XMM-Newton} reveals significant 
evidence for thermal emission from a disk for a TDE.  


This paper is laid out as follows. In \S 2 we discuss our observations and data reduction procedures.  In \S 3 we discuss the results, while in \S 4 we present a discussion of the overall implications of these findings.
 
\section{Observations and Data Reduction}

\begin{center}
\begin{table*}[t]
\caption{VLA Observations}
{\footnotesize
\begin{tabular}{llrcccccccc}
\hline
\hline
Date$^a$& Project$^b$ & $\Delta$ T$^c$ &Config$^d$ &Freq$^e$ (Band)&$\Delta$ B$^f$  & Beam$^g$ & RMS$^h$  & Nuclear Flux$^i$  \\  
&&(days)&&(GHz)&(MHz)&($\prime\prime$, $\prime\prime$, $\circ$) & (mJy/beam) & Density (mJy) \\
\hline
1995-Feb-27 & NVSS$^j$ &  -6150 &D &1.4 (L)& 50&  45, 45, 0&0.45&26$\,\pm\,4$\\
1998-Jul-26 & AB0879$^k$ & -4905 & B & 1.4 (L) & 100   &  5.1, 4.2, 24.0& 1.0   & 34$\pm\,2$\\
2011-Dec-19 & 10C-119$^l$ & $-$11& D & 6.0 (C) & 2048  &  11.0, 9.1, -1.4&  0.015  & 425$\,\pm\,3$\\
2011-Dec-30 (T1)& 10C-119$^l$ & 0    & D & 1.6 (L) & 500  &  38.6, 34.3, -5.22&  0.040 & 211$\,\pm\,3$\\
2012-Feb-24 & 10C-119$^l$ &56& C & 6.0 (C) & 2048  &  3.1, 2.8, -11.7&  0.0039 & 355$\,\pm\,2$\\
2012-Mar-30 & 10C-119$^l$ & 91& C & 1.6 (L) & 500  &  12.2, 11.1, -41.8&  0.045 &  241$\,\pm\,3$\\
2012-Jun-11 & 10C-119$^l$ & 164& B & 1.6 (L) & 500  &  3.5, 3.3, 22.7&  0.018 &  219$\,\pm\,4$\\
2015-Jun-22 & 15A-357$^m$ & 1270 & A & 1.5 (L) & 1000 & 1.73, 1.06, 43.2& 0.12& 115$\,\pm$2 &\\
2015-Jun-22 & 15A-357$^m$ &1270&A &5.5 (C)&1000 & 0.47, 0.33, 46.2&  0.013 &36.3$\,\pm\,0.3$\\
2015-Jul-06 & 15A-400$^n$& 1284&A&9 (X)&2048&0.21, 0.2, 20 & 0.010  & $22.7\pm0.3$  \\
2015-Jul-14 & 15A-400$^n$& 1292&A&9 (X)&2048 & 0.27, 0.16, 43 & 0.020  & $17.1\pm0.2$ \\
2015-Jul-28 & 15A-400$^n$ & 1306 &A&6 (C)&2048& 0.37, 0.28, 48.5 & 0.015&  $33.6\pm0.2$ \\
 2016-Mar-16 & 16A-420$^o$ & 1538 &  C & 6.0 (C)& 2048  &3.5, 3.0, -52.5&  0.012  &   27$\,\pm\,2$&\\
 2016-Mar-16 & 16A-420$^o$ & 1538 &  C & 3.0 (S)& 2048  &6.4, 5.2, -18.4&   0.032 &  $60\,\pm\,3$&\\
 2016-Mar-16 & 16A-420$^o$ & 1538 &  C & 1.5 (L)& 1024 &12.9, 9.9, -21.0&  0.48   &  171$\,\pm\,5$&\\
 2016-May-21 & 16A-420$^o$ & 1604 &  B & 6.0 (C) & 2048  & 0.98, 0.86, -23.5 &   0.034&   $61\,\pm\,2$&\\
 2016-May-21 & 16A-420$^o$ & 1604 &  B & 3.0 (S) & 2048  & 1.8, 1.7, -20.0&   0.058 &   $167\,\pm\,2$&\\
 2016-May-21 & 16A-420$^o$ & 1604 &  B & 1.5 (L)& 1024 &3.6, 3.1, -20.7&  0.50   &  261$\,\pm\,5$&\\
 2019-Apr-21 &  VLASS$^p$ & 2669 &  B &  3.0 (S)&  2048 & 2.9, 2.1, 31.1 &   0.169 & 21$\,\pm\,2$&\\
 2019-Jun-24 &  19A-425$^q$ & 2733 &  B &  1.5 (L)&  64  & 4.9, 3.3, -13.8 &   0.170  & 50$\,\pm\,5$  &&\\
\hline
\end{tabular}}\\
$^a$ Date of observation.\\
$^b$ Project name or code.\\
$^c$ Elapsed time since T1 = Dec 30, 2011.\\
$^d$ VLA array configuration.\\
$^e$ Central frequency of the band followed by the name of the band in parentheses.\\
$^f$ Total bandwidth.\\
$^g$ Synthesized beam major axis, minor axis, and position angle.\\
$^h$ Measured rms map noise (primary beam corrected, near source).\\
$^i$ Flux of nucleus after correction for surrounding disk emission for low resolution images (see text) or directly from Gaussian fitting for high resolution data.\\
$^{j}$ Downloaded and measured from the NRAO VLA Sky Survey (NVSS) \citep{Condon98}. \\
$^{k}$This work. Imaged from the VLA archive. \\
$^l$ \citet{Irwin15}.\\ 
$^m$ Data from \citet{Perlman17}. \\
$^n$ This work.\\
$^o$ This work.  At S-band, the correction for disk emission has been interpolated between known C-band and L-band values.\\ 
$^p$ This work. Measured from VLASS (VLA Sky Survey) data \citep[VLASS;][]{Lacy19}, downloaded from http://cutouts.cirada.ca/ on April 1, 2021.\\
$^q$ This work. Imaged from VLA archive. The uncertainty includes a primary beam error of 5\% for this source \citep{Bhatnagar13}, which is 18.2 arcmin from the field center.
\label{tab:obserinf}
\end{table*}
\end{center}


\subsection{{\sl VLA} Observations}

We present a number of VLA observations of NGC~4845 (Table~\ref{tab:obserinf}), collected here for the first time.  Some are new data that we have obtained, some are re-imaged from previous data in the VLA archive, and some are reduced images from previous surveys. 
Together with previously known values, we now have 20 data points from 15 separate dates.  This fairly extensive lightcurve is one of the most extensive and long-lasting of any TDE.  The only one with similar numbers of long-term points, extending for several years is that of Sw J1644+57, which has more data points but does not extend for quite as many years \citep{Alexander20}.

In Table~1, the first two rows correspond to historical observations of which we include here as a check on the pre-TDE flux level. Following these in the table are the first four epochs of data for NGC~4845 corresponding to the actual TDE event (including epoch T1). These observations were taken in 2011 and 2012 as a part of the CHANG-ES project  \citep[CHANG-ES, observation ID: 10C-119, PI: J. Irwin;][]{Irwin12}. These include L-band and C-band (centering at 1.6 and 6.0 GHz respectively) data in three array configurations, B, C and D. The data reductions are described in \citet{Irwin15}. In 2015, the source was observed in L and C band as part of VLA project 15A-357 (PI: E. Meyer). These data were previously described and published in \cite{Perlman17}.  

The remaining entries in the table correspond to additional subsequent observations of NGC~4845, observed in multiple array configurations and bands (L, S, C and X). 
The separate observations all share a number of steps in reduction procedure, which we describe here. 
We have attempted to measure each new data point in a consistent manner; however an unusual number of co-authors were involved in individually reducing different data sets. Minor differences in e.g., settings during de-convolution imaging and the co-author responsible for the analysis are noted below and are not expected to impact any of the results.

\subsubsection{General VLA data processing}\label{sec:VLAreductions}
All newly presented observations were reduced using the Common Astronomy Software Applications package \citep[CASA;][]{McMullin07} and standard calibration procedures, see e.g. \citet{Irwin13} and \citet{Wiegert15}. 
In short, the data were Hanning-smoothed in order to reduce the effects of strong radio-frequency interference and flagged. We used the primary 3C286 and secondary calibrators (the latter differs between datasets) to correct the antenna baselines, the gain curve, and the antenna based delays. This was followed by additional flagging and gain and bandpass calibrations. The flux density scale was set for each channel in the band using the most recent model for Stokes I \citep{Perley13}. Gain calibrations were performed in two steps, first phase-only calibration, followed by the amplitude calibration. All calibrations were performed twice, with additional flags applied after the first run. 

Imaging deconvolution was done using the CASA task {\texttt{clean}} where we used the multi-scale multifrequency synthesis mode \citep[{\texttt{ms-mfs}};][]{Rau11} to better represent faint emission, with a Briggs robust parameter of 0 (or in some cases 0.5). In general we performed one to two self-calibration rounds for each data set. 

In general, the  somewhat different parameters and other details adopted in the various data reduction procedures described here should have essentially no impact on the reported flux values. The Briggs parameter, which is mentioned in particular, determines the weights of the visibilities when imaging deconvolution is performed. The range of values is from -2 (‘uniform’ weighting) and +2 (’natural’ weighting). The difference between these is basically a tradeoff between better RMS (i.e., sensitivity to faint point sources) versus preserving a higher angular resolution. Natural weighting preserves sensitivity at the expense of resolution, while uniform weighting (which gives more weight to longer baselines) will enhance/preserve angular resolution at the expense of a noisier image.  Values of 0 or 0.5 are extremely common choices representing a compromise between these extremes. Since the flux of our object is well above the noise level, the differences induced by the choice of Briggs parameter is negligible.

The CASA {\texttt{widebandpbcor}} task was used to carry out wide-band primary beam corrections. Flux measurements were made from the primary beam-corrected images in all cases but one. The exception is the the July 2015 data (15A-400), but with the source in the field center (as is the case here), there should be no difference. 

The low-resolution 2016 March and May fluxes at L and C bands, as well as the NVSS measurement, have been corrected for the disk emission of NGC~4845 as discussed in \citet{Irwin15}, subtracting a disk flux of $19 \pm 4$ mJy at L band and $7 \pm 1$ mJy at C band. These disk values were interpolated for S-band. Note that there is no need to do this for the 2015 July C-band data or any of the X-band data due to their higher angular resolution. At high resolution, the fluxes were measured by a Gaussian fit to the central point source using the Gaussian fit tool in CASA.  In cases in which some blending occurred between the central strong peak and the surrounding disk emission, the Gaussian fit was carried out within a region approximately twice the full-width-half-maximum (FWHM) of the synthesized beam. 


\subsubsection{2015 high-resolution observations (ID: 15A-400)}
\noindent
We observed the nuclear region of NGC~4845 with the VLA in A-configuration, in three epochs (2015 July 6, 14, and 28) with a dual-polarization setup. The first two epochs were observed at X-band (centered at 9 GHz), and the third at C-band (centered at 6 GHz). We used J1229+0203 as the secondary calibrator and version 4.5 of CASA for the reductions. 
For the final imaging we used only one combined amplitude and phase self-calibration.

\subsubsection{2016 Swift-concurrent observations (ID: 16A-420)}
\noindent
These observations were taken simultaneously with {\sl Swift} observations on March 16 and May 21, 2016, as a spin-off project of CHANG-ES.  The data were observed at L-band, S-band and C-band (centering at 1.5 GHz, 3 GHz and 6 GHz respectively). We processed the data using version 5.4.0 of CASA and used J1224+0330 as the secondary calibrator in all three bands. 

In addition to manually processing the March observations, we fed them through the VLA pipeline and found the results to be comparable. The May observations were output entirely from the pipeline and double checked before imaging. During imaging, we performed two self calibrations for each data set; in C-band two phase-only calibrations (no improvements could be made after that) and in S-band and L-band one phase-only followed by one amplitude and phase calibration.  

%


\subsubsection{2019 June L-band Observation (ID: 19A-425)}
\noindent
NGC 4845 was observed serendipitously in 2019 June during observations of field J125912+013051 for project 19A-425. NGC~4845 is clearly detected though the source is 18$'$ from the primary beam center, so sensitivity is reduced and the error on the flux is accordingly higher. The data were reduced using the VLA pipeline in CASA (version 5.6.2-3), where J1254+1141 served as the initial phase calibrator. No additional flagging of the data was required and self-calibration was used after initial imaging, with one round of phase-only followed by a final round of amplitude and phase calibration. During imaging, we used a Briggs weighting with robust parameter of 0.5.

\subsubsection{Archival Data from project AB0879}
\noindent
To verify the pre-TDE flux level, we re-analyzed observations of NGC~4845 from 1998 July originally taken for the FIRST 1.4 GHz survey. We imaged field 12585+01310 in which NGC~4845 is 8’ from the primary beam center (at this distance the sensitivity for the pre-upgrade VLA is reduced approximately 12\%). The \texttt{uvfits} data were imported to a CASA measurement set using the task \texttt{importvla} and otherwise standard calibration was applied, with 1354-021 as initial phase calibrator, though the field was ultimately self-calibrated. The final image was made with Briggs robust parameter of 0.5.

\subsection{X-ray Observations}

The X-ray data comes from the Neil Gehrels {\sl Swift} Observatory ({\sl Swift} hereafter)  and {\sl XMM-Newton}  telescopes with detailed information in Table 2. Some early {\it XMM-Newton}  
and {\sl Swift} observations were published in \citet{Nikolajuk13} but are reanalyzed here. The 2016 March and 2016 May observations were meant to coincide closely with two of our {\sl VLA} observations.
Before calibration, we combined
any two Swift/XRT observations that are only minutes or hours apart because sudden gamma-ray bursts can divide a pre-planned Swift observation into two. For XMM, we only keep data from EPIC because it is the only instrument onboard {\sl XMM-Newton} covering soft X-rays with 2D imaging. Then, we use the \texttt{xrtpipeline} command for Swift and the \texttt{emchain/epchain} command in SAS (Science Analysis System) respectively for MOS/PN CCD data of XMM/EPIC to conduct necessary calibrations including calibrating bias, removing bad pixels, removing flares, etc.


\begin{table*}[t]
\begin{center}
\caption{X-ray Observations}
\begin{tabular}{|c|c|c|c|c|c|}
\cline{1-6}
Telescope             & OBSID     & Start Date/ Time    & Total exposure time/s                    & Annotation   & Phase              \\ \cline{1-6} 
{\sl Swift}/XRT             & 31911001  & 2011-01-13  00:03:18.8       & 74.92                                    & Combined  & Bright                 \\ \cline{3-4} 
                      &           & 2011-01-13  00:05:49.5     & 1440.93                                  &                           & \\ \cline{3-4} 
                      &           & 2011-01-13  01:48:20.9       & 1528.33                                  &                         &   \\ \cline{2-4} 
                      & 31911002  & 2011-01-13  03:29:18.9      & 1053.85                                  &                          &  \\ \cline{3-4} 
                      &           & 2011-01-13  05:10:16.7     & 1056.35                                  &                          &  \\ \cline{2-6} 
                      & 31911003  &2012-06-29  07:59:53.3      & 1675.68                                  &Combined    & Faint                 \\ \cline{3-4} 
		     &           & 2012-06-29  11:07:39.4    & 1690.66                                  &                           & \\ \cline{3-4} 
                      &           & 2012-06-29  12:44:31.2      & 1638.21                                  &        &                    \\ \cline{2-6} 
                      & 31911004  & 2016-03-16  05:11:55.2    & 177.31                                   & Combined     & Faint              \\ \cline{3-4} 
                      &           &2016-03-16  07:02:04.8      & 169.82                                   &          &                  \\ \cline{3-4} 
                      &           & 2016-03-16  07:02:04.8      & 177.31                                   &              &              \\ \cline{2-4} 
                      & 31911005  & 2016-03-16  05:15:15.6       & 1293.59                                  &            &                \\ \cline{3-4} 
                      &           & 2016-03-16  07:05:16.2   & 1233.66& &\\ \cline{3-4} 
                      &           & 2016-03-16  08:26:16.0     & 1293.59&&  \\ \cline{2-6} 
                      & 31911006  & 2016-05-21  03:32:31.0     & 202.28                                   & Combined    & Faint               \\ \cline{3-4} 
                      &           & 2016-05-21  05:09:10.2     & 162.32 & &\\ \cline{3-4}                
0                      &           & 2016-05-21  06:48:06.6  & 167.32                                   &                &            \\ \cline{2-4} 
                      & 31911007  & 2016-05-21  03:36:15.8      & 1113.79                                  &                         & \\ \cline{3-4} 
                      &           & 2016-05-21  05:12:16.3      & 1473.4                                   &                          &  \\ \cline{3-4} 
                      &           & 2016-05-21  06:51:15.6    & 1233.66                                  &                           & \\ \cline{2-6} 
                      & 31911008  & 2017-11-12  00:08:34.7       & 1333.55                                  &      Combined   & Faint                   \\ \cline{3-4} 
                      &           & 2017-11-12  01:53:17.4      & 991.42                                   &                    &       \\ \cline{1-6} 
{\sl XMM-Newton}            & 658400501 & 2011-01-22 13:42:32& 8529                  & EPIC data & Bright\\  
				&			&			&			&		not included & \\ \cline{2-6} 
                      &  &  & MOS1:15165.5 & &   \\
                      &		658400601	   &     2011-01-22 16:23:28                  & MOS2:16632.0&  & Bright \\ 
                       &			   &               & PN:9645.6 	&  &                       \\ \cline{1-6} 
\end{tabular}
\end{center}
\label{SwiftTable}
\end{table*}

With {\sl Swift} and {\sl XMM-Newton} event files that were cleaned of background flares, we choose source and background radii according to the central PSFs of the telescopes, and extracted source and background counts using typical procedures{\footnote{For {\sl XMM-Newton} details, see \citet{Weaver18}, while for {\sl Swift}, see https://swift.gsfc.nasa.gov/analysis/}}.
We generated Swift response files by selecting the correct RMFs in its calibration database based on their observation times and calculate ARFs by “xrtmkarf” command. 
For {\sl XMM-Newton}, the “rmfgen” and “arfgen” commands in SAS are used to produce all response files. 
We use absorbed power law plus absorbed APEC models, according to the expression ${\rm {TBabs_G (APEC+TBabs(power law))}}$, where ${\rm TBabs_G}$    is the Galactic absorption,
    {\footnote{The APEC model is discussed in detail at the website http://www.atomdb.org. The atomic database AtomDB includes the Astrophysical Plasma Emission Database (APED) and the spectral models output from the Astrophysical Plasma Emission Code (APEC). The APED files contain information such as wavelengths, radiative transition rates, and electron collisional excitation rate coefficients. APEC uses these data to calculate plasma model spectra. The APEC output models in AtomDB are for optically-thin plasmas in collisional ionization equilibrium.}}  
for the two bright-phase observations in Table 3, accounting for both the nucleus and the surrounding hot gas.  We considered the source to be extincted by the Galactic column of $N_{\rm H}^{\rm Gal}=1.67 \times 10^{20}~{\rm cm^{-2}}$ \citep{Dickey90,Stark92}, and absorption from the host galaxy was allowed to vary freely.  The default version of the solar abundances were assumed for all APEC models. All spectra are regrouped to contain at least 30 counts per bin, which should be sufficient for using a simple $\chi^2$ minimization.  Although the C-statistic is statistically more accurate, we have not used it because its use would involve complication and uncertainty in having to model the background spectral contribution.



\section{Results}

In this section we discuss the results of those observations as well as those in the X-rays. Table 1 contains the radio data and Table 2 contains the x-ray data.

\begin{figure*}[ht]
\center{\includegraphics[width=1.\linewidth]{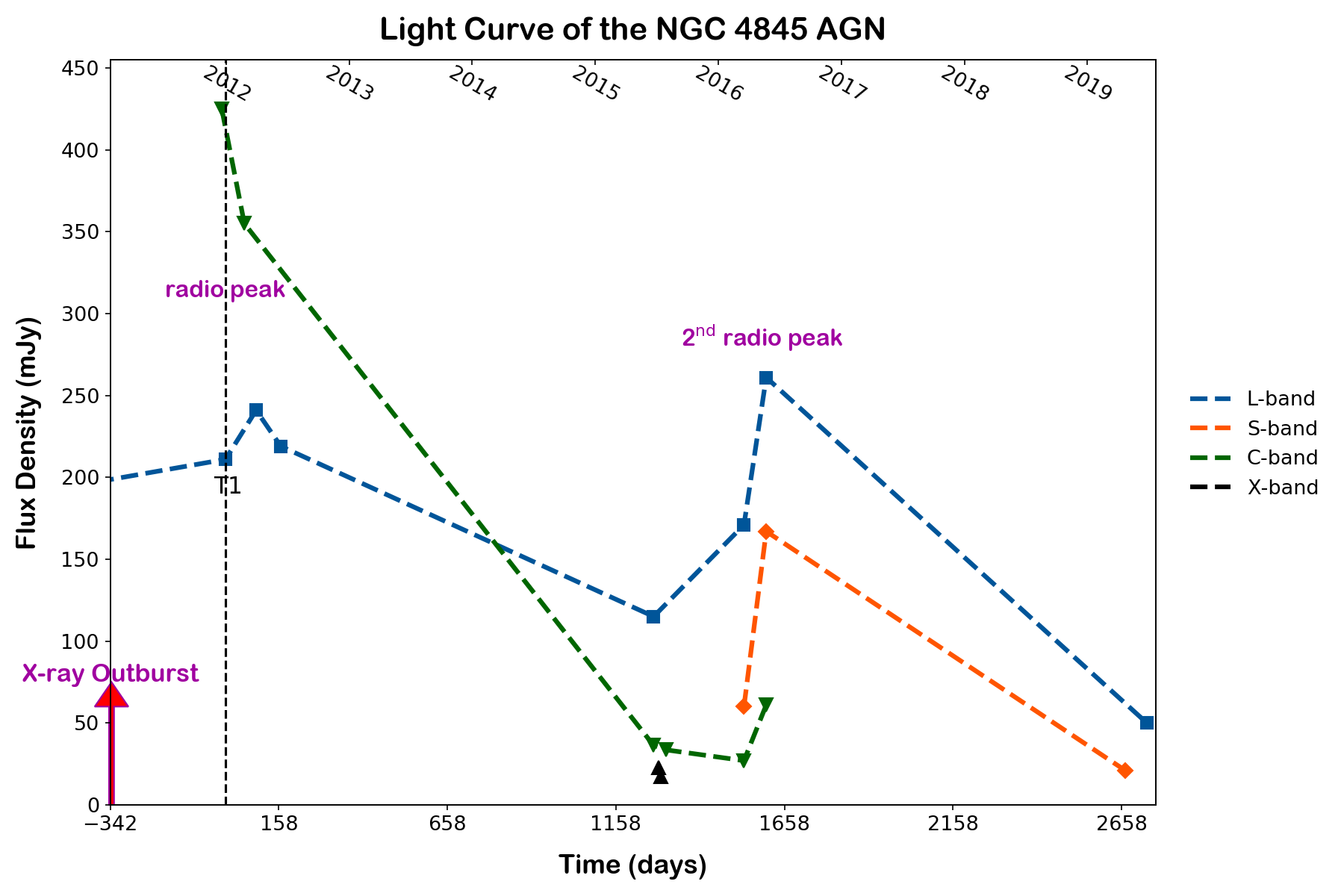}}
\caption{Nuclear flux density from Table~1 as a function of time, measured from T1 (T = 0). Error bars are plotted but typically have sizes that are smaller than the icons. Two very early points at L-band have been omitted for ease of viewing. The X-ray outburst (denoted as T0$=-342$ days=2011-01-22, as consistent with \cite{Irwin15} and other papers}) is marked at far left.  The first and second radio peaks are marked. L-band is shown in blue, S-band in red, C-band in green, and X-band in black.
\label{fig:radioflux}
\end{figure*}

\subsection{Continued evolution of the Nuclear radio flux}

Consistent with \citet{Irwin15}, we define 2011 Dec 30 as time T1 ($t=0$ on the plot), and plot the other data given here and in \citet{Perlman17} in Figure 1. 
As can be seen in Figure 1, the nuclear radio flux declined monotonically from the radio peak until approximately 2015.  In 2015, we saw a plateau in the radio flux, with  20\% temporal variability in the X-band between 2015 July 06 and 2015 July 14 (the apparent C-band variability seen in June-July 2015 is more likely a result of the two slightly different radio frequencies). Then, in 2016 (day 1538), a sharp increase began in the radio flux.  By day 1604 the flux in C, S and L band had increased by a factor of 2.5-3.  This defines the epoch of our new `flare'.
Unfortunately there is almost no information on the radio variability of the nuclear source of NGC 4845 
previous to the TDE, although we note that the two 
L-band fluxes in 1995 and 1998 are consistent with 
one another at the 2 $\sigma$ level.


\subsection{Results from the reanalysis of {\sl Swift} and {\sl XMM-Newton} data}


Power-law fits to 
both {\sl XMM-Newton} and {\sl Swift} data (Nikolajuk \& Walter 2013) found a steep  ($\Gamma \approx 2.35$)
X-ray spectrum with a column density $N_H \approx 7 \times 10^{22}~ {\rm cm^{-2}}$.  Our reanalysis confirms that result, but we find that an APEC model is a better fit to the data, with the main improvement being an improved fitting of the soft excess. A simultaneous fit of all three 2011 datasets yielded a 
power-law $\Gamma = 2.316 \pm 0.042$,  $kT = 0.238_{-0.052}^{+0.149}$ keV for the thermal component, and absorbing
column $N_H = 8.8 \pm 0.2 \times 10^{21} ~{\rm cm^{-2}}$.  The decrease in the $\chi^2_\nu$ was highly significant -- from 0.973 to 0.410 for the {\it Swift} data and from 1.291 to 0.710 for the {\sl XMM-Newton} observations.  The soft X-ray component is also noted by Nikolajuk \& Walter (2013), but
those authors suggest it is unrelated to the nuclear emission associated with the TDE based on a fit to a blackbody component with temperature $kT=0.33 \pm 0.04$ keV  that is extincted only by the Milky Way's column, which they claim (their section 3.2) is consistent with diffuse emission from galaxies (Bogd\'an \& Gilfanov 2011; Jia et al. 2012).  They do not provide any statistics on that fit; however, it would be one additional degree of freedom beyond our model, which as noted shows a large improvement over a power-law.   We suggest that the evidence for a smaller column for the thermal component is not persuasive, and moreover we point out that diffuse emission from a galaxy at only 17 Mpc distance would easily be resolved with {\sl XMM-Newton}.  Moreover, the 0.5-7.0 keV flux from the thermal component, $4.3\times10^{-13} {\rm ~erg~cm^{-2}~s^{-1}}$, differs by more than 3$\sigma$ from the value observed in 20160-2017  (Table 3). We therefore assess that the soft component is more likely to be nuclear in origin.

Table 3 presents the X-ray source and background region information, as well as counts and fluxes associated with each observation.  The fluxes for the 2011 {\sl XMM-Newton} and {\sl Swift} observations come from the procedure discussed above. For the 2012-2017 Swift observations, the count rates are far too small to allow for any spectral fitting. We thus use the above best-fit spectral model to convert the count rates to the corresponding energy fluxes for comparison.    For the two bright phase (2011) observations plus the 2012 {\sl Swift } observation, we were able to extract background information from the individual observations. However, for the 2016-2017 {\sl Swift} observations, where this was not possible with the individual observations, the background levels were taken from the combined 2016-2017 observations. 
Figure \ref{fig:XrayFlux} shows the 0.5-7 keV X-ray fluxes in the 0.5-7 keV band (with no absorption correction, thus allowing us to include the INTEGRAL observation) as a function of time.
In addition, we show in Figure \ref{fig:XrayFlux} the X-ray fluxes from \citet{Nikolajuk13}, extrapolated down to the 0.5-7 keV band using the joint {\sl INTEGRAL} + {\sl XMM-Newton} best-fit model in that paper, namely a hard X-ray spectral index of $\Gamma=2.22 \pm 0.03$, slightly different from what we find here.

Three things are clear from Figure \ref{fig:XrayFlux}. First, IGRJ12580+0134 represented a much larger departure from the ``quiescent'' level of X-ray flux than was previously realized. Based on the {\sl INTEGRAL} lightcurve of \citet{Nikolajuk13} one could only say that the event was an increase in X-ray flux of at most a factor 10. Figure \ref{fig:XrayFlux} makes it clear that IGRJ12580+0134 represented a much larger increase in NGC 4845's X-ray flux -- more than two orders of magnitude.
Second, as also found by \citet{Nikolajuk13}, there is no evidence for spectral curvature between 7 keV and the 17.3-80 keV band of INTEGRAL.  Not only are the power-law components of the {\sl INTEGRAL} spectral fit and our {\sl XMM-Newton + Swift} fit roughly consistent within the errors (they differ by less than 2 $\sigma$), but in addition, the 2011 {\sl Swift/XRT } and {\sl XMM-Newton} fluxes fit smoothly on the curve described by the {\sl INTEGRAL} points, without any evidence of error due to extrapolation.  
Finally, there is no evidence that the 2016 radio flare described in \S 3.1 (Figure 1) was accompanied by an X-ray flare. While the two 2016 data points are not detections (and for them as well as the 2017 Swift observation we show 2$\sigma$ upper limits), we see that when the three 2016-2017 observations are added together a significant detection is seen and the detected flux is within $1\sigma$ of the 2012 data point.  If indeed there had been an X-ray component to the 2016 flare, the two 2016 {\sl Swift} observations would dominate that point, and we see no evidence of that.

\begin{table*}[ht]
\begin{center}
\caption {X-ray Source and Background Details}
\begin{tabular}{|c|c|c|c|c|c|c|c|}
\hline
Observation(s)                                      & PSF & Center  & Source & Background & Source & Background & 0.5-7.0 keV Flux  \\
							   &           &              & Radius	& 	Radius & Counts & Counts & ${\rm (erg~cm^{-2}~s^{-1}})$\\ \hline
00031911001 + 00031911002 &       &                    &                        &                       & 2026                                & 283                              & $ 2.60\pm 0.06 \times 10^{-11} $           \\ \cline{1-1} \cline{6-8} 
00031911003                          &       & DEC=               &                        &                            & 9                                   & 4                                & $2.4^{+1.3}_{-0.9} \times 10^{-13}$             \\ \cline{1-1} \cline{6-8} 
00031911004 + 00031911005  & 18\farcs   & +1d34m33s,         & 30\farcs                     &        30\farcs-60\farcs                    & 6                                   & 1                                & $<3.75 \times 10^{-13}$           \\ \cline{1-1} \cline{6-8} 
00031911006 + 00031911007  &       &                    &                        &                            & 6                                   & 2                                & $<3.75 \times 10^{-13}$            \\ \cline{1-1} \cline{6-8} 
00031911008                          &       &                &                        &                            & 6                                   & 3                                & $<7.35 \times 10^{-13}$ \\\cline{1-1} \cline{6-8} 
00031911004-00031911008                      &       & RA=                &                        &                            & 18                                   & 6                               & $1.8^{+0.7}_{-0.5}\times 10^{-13}$ \\ \cline{1-2} \cline{4-8} 
                             &      & 12h58m1s.3         &                     &                     & MOS1: 14091  & MOS 1: 9439 & \\  \cline{6-7}
0658400601&			6\farcs					    &        & 10\farcs &10\farcs-30\farcs &  MOS2:14930 & MOS2: 9902 & $2.87^{+0.08}_{-0.03}\times 10^{-11}$ \\  \cline{6-7}
&								    &         & & &  PN: 28475 & PN: 16216 & \\  \hline
\end{tabular}
\end{center}
\label{Xraydetails}
\end{table*}

\begin{figure} 
\center{\includegraphics[width=1.09\linewidth]{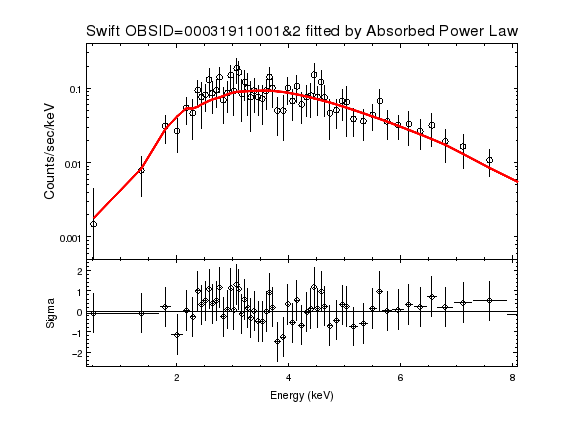}
\includegraphics[width=1\linewidth]{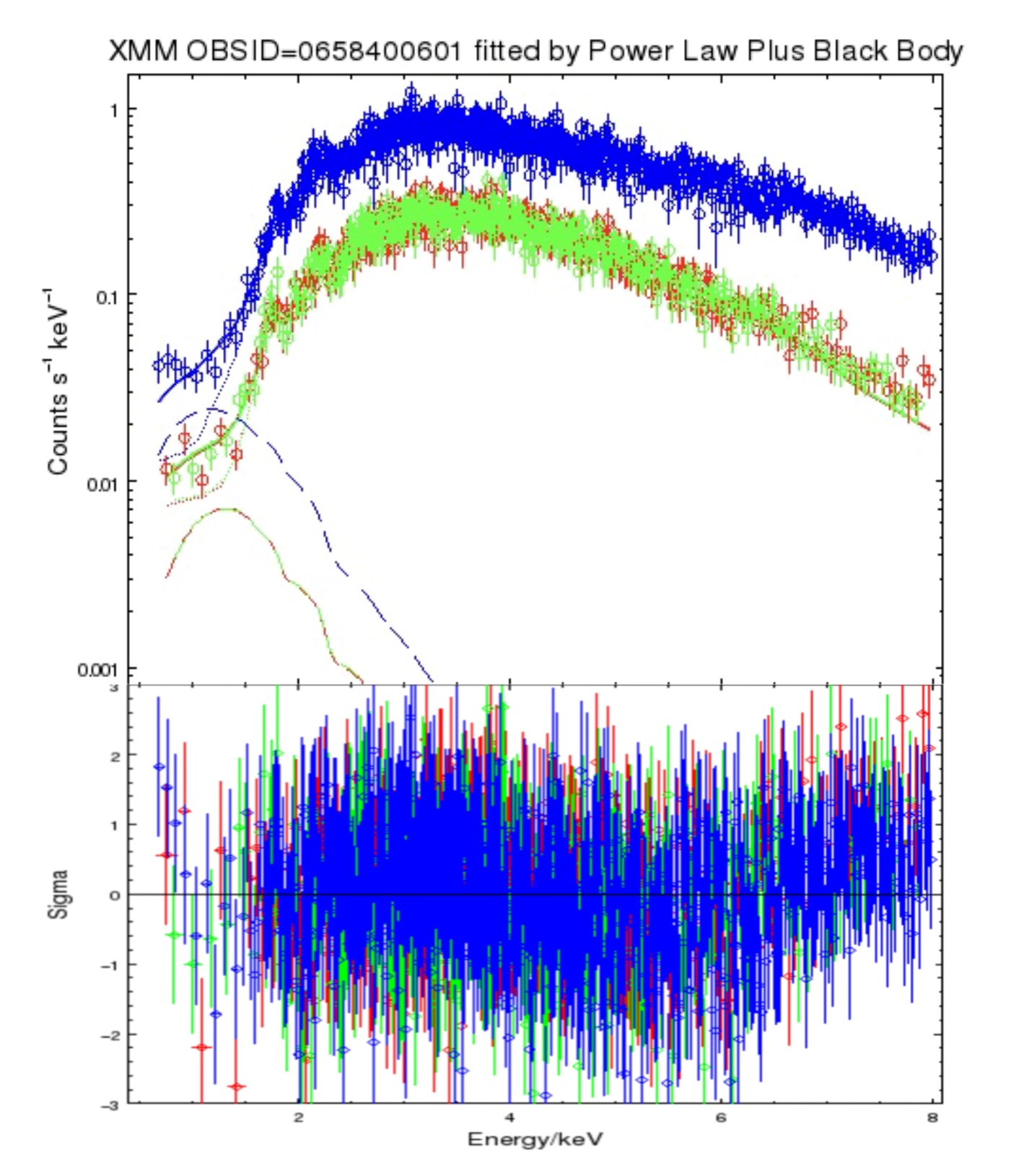}}
\caption{X-ray data for IGRJ12580+0134, with {\sl Swift} data shown at top and {\it XMM-Newton} data shown at bottom.  The 
residuals shown in each panel are for the spectral models discussed in \S 3.2.}
\label{fig:XraySpec}
\end{figure}

\begin{figure}
\center{\includegraphics[width=1.05\linewidth]{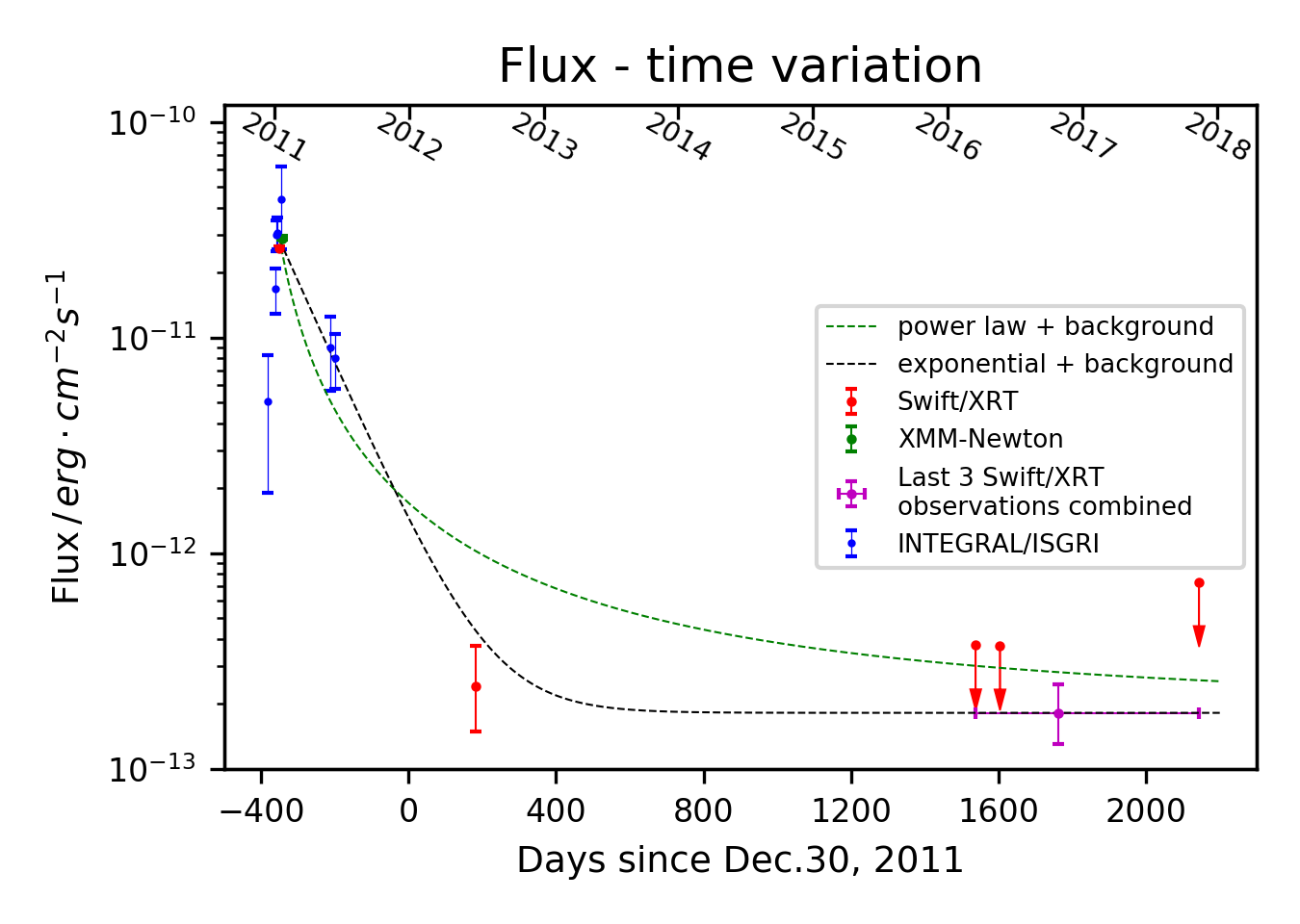}}
\caption{X-ray fluxes for IGRJ12580+0134 plotted as a function of time.  Two curves are overplotted:  In black, an exponential + a constant background level, and in green, a $t^{-5/3}$ power law plus exponential.  See \S 3.2 for details.}
\label{fig:XrayFlux}
\end{figure}

%

\section{Discussion}

\subsection{Classification \& X-ray Spectrum of IGRJ12580+0134}

Recently, \citet{Auchettl17} have questioned the classification of IGR J12580+0134 as a TDE, 
based on the fact that the host galaxy, NGC 4845, is a Sey 2/LINER \citep{Ho95}. They claimed that IGRJ12580+0134 could be explained by AGN processes{\footnote{The source is listed in the ``TDE Graveyard'' section of the Open TDE Catalog, https://tde.space, presumably for similar reasons.}}. 
The nucleus likely has some X-ray variability, so we definitely agree with the general cautions regarding AGN variability expressed by those authors.
But we contend that IGR J12580+0134 is still likely to be a TDE, for several reasons.  

First, the large X-ray and radio flux increases
observed \citep[now seen to be at least two orders of magnitude in X-rays, much larger than could be justified based on the {\sl INTEGRAL} data in][]{Nikolajuk13} are well beyond what is usual for AGN, with less than 1\% of AGN having flares of that magnitude \citep{Auchettl17}. The flares that tend to be similarly extreme are in blazars \citep[e.g., the 1998 flare of PKS 2005-489;][]{Perlman99} and narrow-line Sey 1 galaxies \citep[e.g., 1ES 1927+654;][]{Trakhtenbrot19a,Trakhtenbrot19b,Boller03}, and NGC 4845 displays no properties in common with such sources.  Also, the nuclear luminosity in quiescence would be $\leq 6.2\times 10^{39} {\rm~erg~s^{-1}} $, as observed in  the 2016-2017 {\it Swift} observations (of which some could be due to stellar or gas processes), much more in line with a quiescent galaxy nucleus or low-luminosity  AGN (e.g., Ptak et al. 1999).  

Second, \citet{Nikolajuk13} pointed out that the hard X-ray (17.3-80 keV) flux evolution after the peak was consistent with a $t^{-5/3}$ law based upon {\sl INTEGRAL} data. Arguments consistent with Kepler's second law
predict that the mass infall rate should 
approach  $t^{-5/3}$ at late times for all types of disrupted stars \citep[][note 
that the evolution close to the peak can depend on the stellar structure]{Phinney89,Lodato09}.
However, if a thermally emitting disk forms, the evolution can be very different, $\propto t^{-5/12}$ if emission sits in the Rayleigh-Jeans tail of a thermal component. Indeed, at much later times a $t^{-5/3}$ evolution is not fully consistent with the X-ray lightcurve. (Figure 3). Alternative fits are also
possible, including an exponential fit and/or power law at earlier time with an exponential decline at later times.  Exponential behavior in the X-ray/optical has been observed in some TDEs, including e.g., iPTF16fnl \citep{Onori19}, another faint TDE.  Such behavior is not unexpected in TDE, as \cite{lodato11} point out, the $t^{-5/3}$ behavior is expected to continue for only a few months before steepening to become essentially exponential in the X-ray band, where the X-ray emission at later times would be dominated by an outflow.

Figure 3 makes clear that the 2012 {\sl Swift} flux fits neatly on that same power-law as the early {\sl INTEGRAL data}. Thus the $t^{-5/3}$ behavior continued for nearly 500 days.  In addition, the quickness of the decline is not consistent with an extinguishing event, which should take place over much longer timescales ($\sim 10^4$ years, Shen 2021).

The X-ray spectrum of IGRJ12580+0134 also has several properties that are more in common with 
TDEs. Its spectral index, $\Gamma=2.316\pm 0.042$, is rare among AGN \citep[which are dominated by objects with $\Gamma \approx 1.8$;][]{Auchettl18,Ricci17} but common among TDEs \citep{Auchettl18}.  While the X-ray power-law emission of IGRJ12580+0134 continues to 80 keV \citep{Nikolajuk13}, unlike any other TDE, the power-law component of at least one other jetted TDE, 
Swift J1644+57,  continues up to at least $\sim 30$ keV \citep{Bloom11,Kara16}, as evidenced by its Swift/BAT detection. 
X-ray spectral indices over 2 are seen in HSP BL Lac objects \citep[e.g.,][]{Perlman05,Paliya19}, 
where the X-ray spectrum is dominated by the tail of synchrotron emission, but IGRJ12580+0134 shows no evidence for beaming. A thermal excess is almost universally found in TDEs, but the temperature, 
$kT = 0.24_{-0.05}^{+0.15}$ keV, is somewhat higher than seen in most TDEs \citep[typically $\sim 50$ eV, although comparably high temperatures are seen in $\sim 10\%$ of cases;][]{Auchettl18} and AGN \citep[typically $\sim 100$ eV but a with a handful over 200 eV;][]{Ricci17}.  As discussed in \S 3.2, this thermal component is very likely to be of nuclear origin.
Finally, the absorbing column is reasonable both for AGN and TDEs \citep{Auchettl18} and is similar to that seen in the jetted TDE Swift J1644+57.

This re-assessment confirms that IGR J12580+0134 was a TDE in at most a low-luminosity AGN. Two very recent reviews make the point that TDE and TDE-like events should be common in AGN \citep{Gezari21,Saxton21}.  \cite{Saxton21} specifically discusses IGR J12580+0134 and mentions this possibility. As Saxton et al. note, at least a few other possible TDEs fall into this category.  One particularly interesting case could even be a recent event in OJ287 where the X-ray lightcurve shows a clear, $t^{-5/3}$ decay after the event \citep{Huang21}. A much more doubtful case is Arp 299, where \citet{Mattila18} suggested the presence of a TDE, attributing the lack of signatures in other bands to a large obscuration. However, that event was more than an order of magnitude smaller in amplitude than IGR J12580 and lacked the characteristic lightcurve variability.
A recent paper by \cite{Zabludoff21} discussed a more sophisticated way to discriminate TDEs from interloper events, which will be critical for future surveys such as the LSST.

The clarification of IGR J12580+0134's nature as a TDE also allows us to discuss other aspects of its X-ray spectral behavior.
The flow of matter into the near vicinity of an active black hole might make TDEs more common in AGN than in quiescent galaxies, although this hypothesis needs to be tested by {\it LSST} observations.  
\citet{Stolc19} have pointed out that in such systems one might expect an evolving, relativistic
Fe K$\alpha$ line.  Our X-ray spectral data do not show such a line, but the signal to noise at 6.4 keV is low.

\subsection{Continuing Radio Flux and Spectrum Variability}

As discussed in \S3, the radio flux of IGR12580+0134 has continued to evolve, as shown in Figure 1.
Up to and including 2015, VLA data show descent to a plateau in the radio flux.   During that plateau, $\sim 20\%$
variability was seen in the X-band, which is common in radio-loud AGN and also more quiescent galaxies. For an 
example of the latter, see the lightcurve of Sgr A* shown in \cite{Subroweit17}.  Very little data exist on the long-term radio lightcurves of TDE \citep{Alexander20}, so it is difficult to comment on the frequency of such small scale radio variability in that population.  
The plateau in the radio emission seen in 2015 is reminiscent of
events seen in the X-ray lightcurve of ASASSN-14li, 1-2 years after the TDE \citep{Bright18}.
In that object, \citet{Pasham18} have detected a time-lag between the soft X-ray and radio emission, and interpreted it as linear disk-jet coupling. Unfortunately, we do not have sufficient radio or X-ray data in IGR J12580+0134 to perform such a test.
In addition, we have found 
a major radio flare in 2016. Our sole radio point after 2016, taken in 2019, shows a large decline in  radio flux, however due to the lack of monitoring between 2016-2019 we cannot speculate on the speed of the decline.    

\begin{figure}
\center{\includegraphics[width=\linewidth]{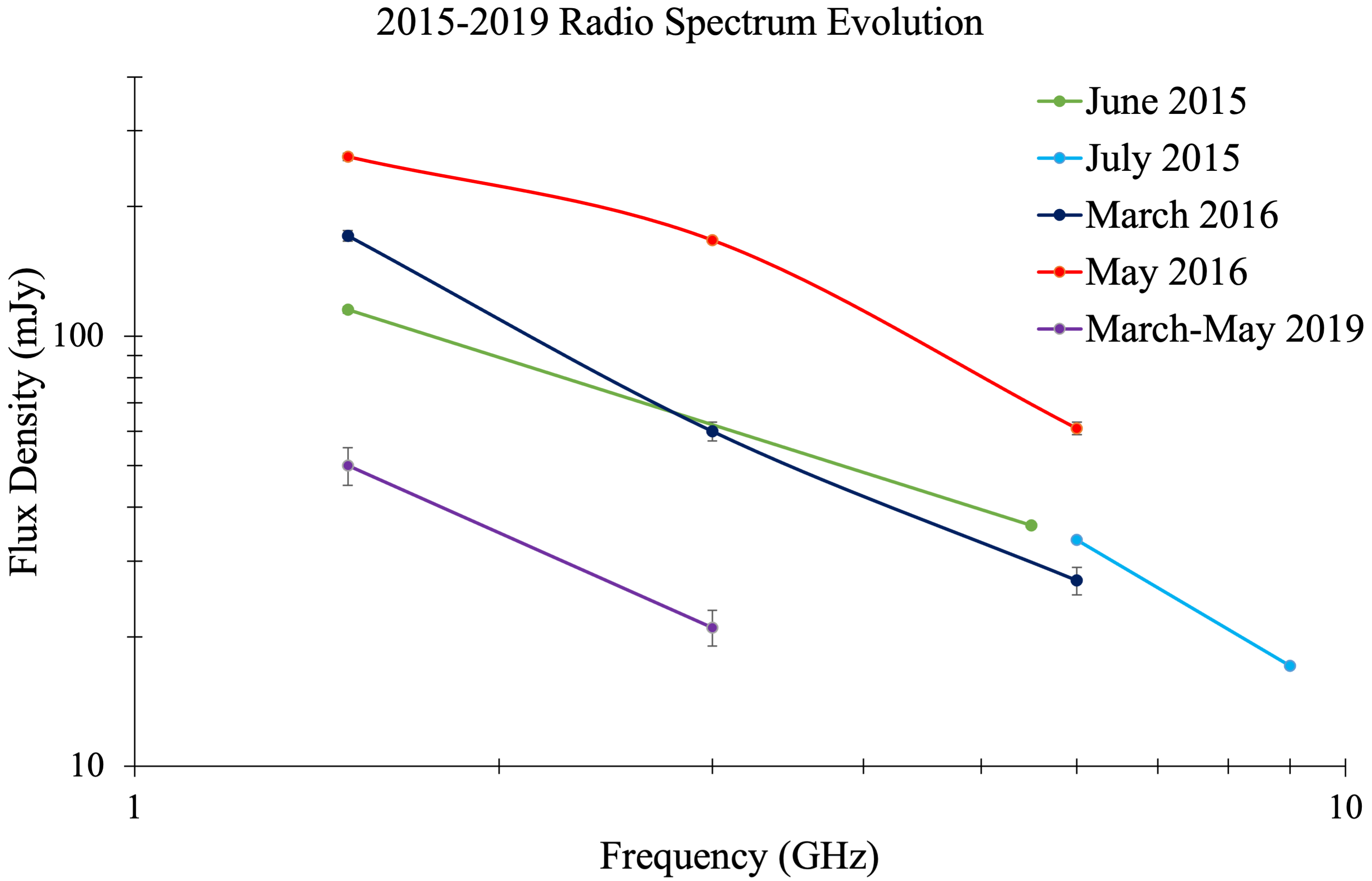}}
\caption{AGN flux density from Table 1 as a function of frequency for the time period 2015-2019. 
Spectra from the time periods 2015 June, 2015 July and 2019 Mar-May are not truly simultaneous and combine all observations from the noted time period.  Both axes are logarithmic, and error bars are shown (Table 1). See \S 4.2 for discussion.}
\label{fig:radiospectrum}
\end{figure}

It is interesting to look at the evolution of the radio spectrum during these events, shown in Figure 4. In that plot, two of the spectra are truly simultaneous (March 2016 and May 2016), while the others simply show all data during the indicated periods. In 2015, we saw a C-L band spectral index of $\alpha_{CL}=0.89$, while the X-C band spectral index in July 2015 was $\alpha_{XC}=1.67$ (the slightly different June and July 2015 C-band flux points is clearly a result of observing at two different central frequencies, as Figure 4 makes clear). The spectral index at lower frequencies (L and S bands) was at its flattest during the flare's brightest stage, in May 2016, when a value of $\alpha_{SL}=0.64 ~(F_\nu \propto \nu^{-\alpha})$ was seen, as compared to a value of $\alpha_{SL}=1.51$ in March 2016 when the flare was in an earlier stage. However,in May 2016, the higher frequency (C to S band) spectral index was significantly steeper ($\alpha_{CS}=1.45$) than it was in March 2016 ($\alpha_{CS}=1.15$).  Thus the radio spectrum in March 2016 steepened moderately as one went to lower frequencies, while in May 2016 it flattened drastically as one went to lower frequencies.  
The new flare has not been monitored sufficiently to show the decline behaviour between 2016-2019. However, in 2019, the radio spectrum between S and L bands was $\alpha_{SL}=1.25$.

IGR J12580+0134 joins ASASSN-15oi \citep{Horesh21} as one of only two TDEs with late-time radio flares observed.  Comparing the two, ASASSN-15oi during the peak of its late-time radio flare was two orders of magnitude fainter in flux than IGRJ12580+0134, but due to its much greater distance (216 Mpc) its peak luminosity was a factor of a few higher, actually higher than its radio peak during the primary TDE event. The increase in radio luminosity observed in ASASSN-15oi was actually larger than that observed in IGRJ12580+0134 -- a full order of magnitude, as compared to a factor 3 here.  The two occur at similarly late times after the TDE (4 years after the TDE in ASASSN-15oi as compared to 5.5 years after the TDE in IGRJ1258+0134).   However,  apparently no radio monitoring of ASASSN-15oi exists after the late-time rise in 2019, at least up until the publication this year of Horesh et al. (2021).

There are to our knowledge four other TDE that show late-time X-ray flares. These are PS10adi \citep{Jiang19}, OGLE16aaa \citep{Kajava20}, ASASSN-15oi (at the same time as its radio flare, \cite{Gezari17,Holoien18}), ASASSN-18jd \citep{Neustadt20}, and AT2019azh \citep{Liu19}. All four have been modeled as possibly due to a second fallback event due to circularization of the orbits of accreting material \citep{vanVelzen19,Gezari21} or a sub-relativistic shockwave \citep{Alexander20, Yalinewich19}. As  discussed below, the radio-only flare in IGR J12580+0134 cannot be explained in this context, similar to ASASSN-15oi where a similar conclusion was reached (Horesh et al. 2021). 

\subsection{Radio flare from a jet-cloud encounter}

The 2016 radio flare seems more likely to be the result of the jet intercepting a circumnuclear cloud and accelerating particles to relativistic energies. Following the modeling of an off-axis jet propagating in the CNM employed to explain the multi-wavelength evolution of radio emission from IGR J12580+0134 \citep{Lei16,Perlman17}, we fit the fluxes assuming that a denser gas cloud exists in the path of the propagating jet. The jet dynamics is described by a set of hydrodynamical equations, including the evolution of the radius and Lorentz factor of the jet, and the swept mass \citep{Huang00,Liu20}. The magnetic field and accelerated electron population share parts of the thermal energy of the downstream medium of the shock, with fractions $\epsilon_B$ and $\epsilon_e$, respectively. Electrons were accelerated to form a power-law spectrum with an index of $-p$. Synchrotron radiation was then calculated given the magnetic field and electron distribution at each time. 

\begin{figure}[htb]
\center{\includegraphics[width=1.05\linewidth]{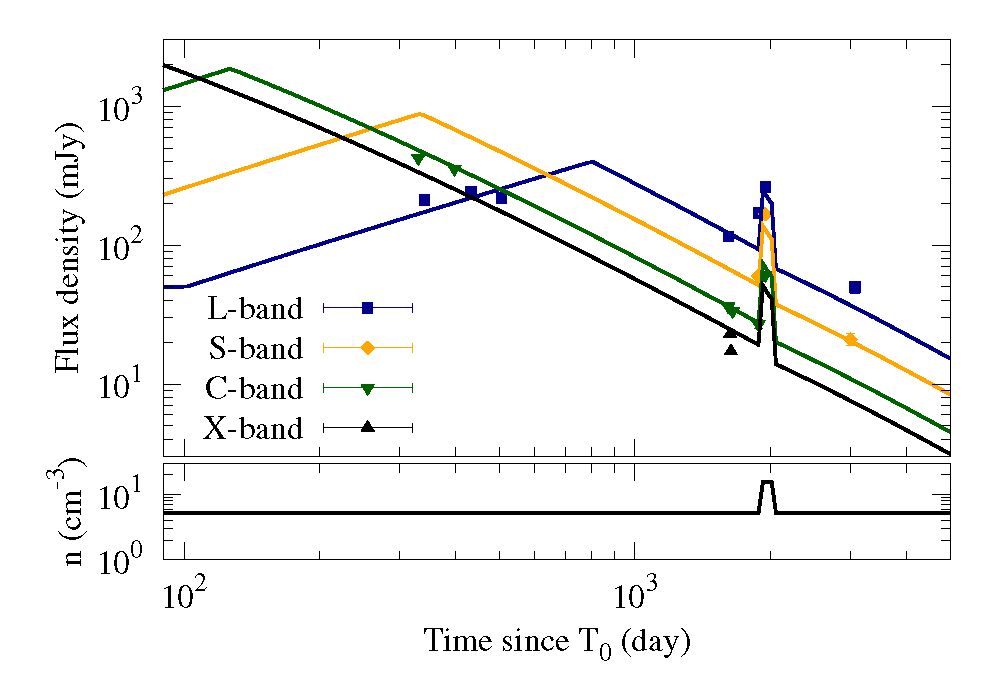}}
\caption{Model fitting of the radio light curves (upper sub-panel) of IGR J12580+0134 in a scenario with a jet propagating in the CNM. The radio flare is explained by the encounter with a gas cloud as indicated by the density profile in the bottom sub-panel.}
\label{fig:model}
\end{figure}

Figure \ref{fig:model} shows the light curves from the model calculation, compared with the data. 
The main model parameters include: the total kinetic energy of the ejecta $E_k=2.5\times10^{52}$ erg, the initial ejecta Lorentz factor $\Gamma_i=8.5$, $\epsilon_e=0.27$, $\epsilon_B=0.09$, the jet opening angle $\theta_j=5.4^{\circ}$, the viewing angle $\theta_{\rm obs}=36^{\circ}$, the electron spectral index $p=2.80$, and the background medium density $n=5.2$ cm$^{-3}$. A gas cloud with an average density of $n_{\rm cl}=10.3$ cm$^{-3}$ spanning from 0.95 to 0.97 pc from the central black hole, as indicated by the bottom panel of Figure \ref{fig:model}, is found to be able to fit the radio flare. The cloud mass swept by the jet (with opening angle of $\theta_j$) is estimated to be about $1.8\times10^{-4}$ M$_{\odot}$, and the jet velocity is about $0.17c$ when it hits the cloud. Given the soft spectrum of electrons and the cooling effect, we estimate that the flux in the X-ray band (at 1 keV) during the radio flare time is about $3.4\times10^{-17}$ erg~cm$^{-2}$~s$^{-1}$~keV$^{-1}$, which is several orders of magnitude lower than the upper limits shown in Fig. \ref{fig:XrayFlux}.

Note that there is degeneracy among the model parameters, and the above parameters are one example of those with reasonable values. Nevertheless, some of these parameters can be constrained by the data. While the CNM density degenerates with the total kinetic energy as well as the energy partition fractions, the density contrast between the gas cloud and the CNM background can be  constrained relatively well (about 2, although the exact value is affected somehow by the thickness of the shell). The swept mass, derived by the gas density and cloud thickness, is constrained by the declining profile and post-flare fluxes. 

This simplified model fits the observational data, including the radio flare, fairly well. The flattening of the radio spectral index at the brightest phase of the late-time radio flare (Figure 4) is consistent with this model, as such an interaction would cause a shock in the jet. The rapid steepening of the radio spectral index in the later 2016 epoch is consistent with a small size for the cloud. The late-time L-band flux in 2019 is a little bit higher than the model expected flux. The X-band fluxes also show a quick decrease which was not observed in other bands. Further refinement of the model may improve the match between the model and data. If this scenario is correct, the radio monitoring of jetted TDE could allow us to probe the surrounding medium of the galactic nuclei. The density profile of the background CNM can in principle be probed with the light curves \citep{Berger12}. It is, however, difficult for IGR J12580+0134 given the sparse data points during and immediately after the flare, since as can be seen additional observations would have been helpful.

Horesh et al. (2021) reached a similar conclusion regarding the nature of the late-time radio flare of ASASSN-15oi.  They modeled it as due to the collision of the radio jet with a circumstellar shell of material, after rejecting fallback events and shocks within a structured jet due to the steepness of the radio rise.   We cannot distinguish that possibility from the model we propose in this paper, and in fact such a shell would be consistent with the possibility we model, although resulting in a considerably greater mass for the disturbance.

\section{Conclusions} 

TDEs are some of the most energetic events that occur in a galaxy nucleus. Recent works have shown the similarities of TDEs to AGNs.  Both are associated with accretion events to the nuclear supermassive black hole -- relatively long-lived in the case of AGN, and temporary and short-lived in the case of TDE.  The short-lived nature of TDE may, in fact, be one reason for the differences between AGN and TDE X-ray spectra.
The latter are often blackbody in nature, and can be explained as material that at any time is orbiting at a small range of radii.  By comparison, the power-law nature of AGN X-ray spectra require matter orbiting at a large range of radii \citep{perl13}, both within the accretion disk and its corona.  This necessarily requires a long-lasting, stable accretion structure with a more continuous accretion stream rather than a discrete event.  In this respect, a late-time radio flares associated with the interaction of the outflow with an ISM cloud, should not be regarded as a surprising -- in fact, radio-only flares in AGN are commonly modeled in such a fashion (e.g., \cite{hervet17} and references therein).  

Other possibilities must be considered for such a radio flare.  The first is the ejection of a new component within the jet very close to the origin of the outflow near the supermassive black hole.  This would be similar to what occurs in blazars and other AGN during large radio flares (e.g., \cite{Larionov20, bland19}). However, such flares usually are associated with multiwavelength events, making such an event unlikely. In addition, a component would become visible at a later epoch, a possibility we discuss in a future paper (Perlman et al., in prep).  A second possibility is shockwaves in a structured jet, also often invoked for blazar variability but rejected by Horesh et al. (2021) due to the lightcurve shape. A third possibility is a delayed ejection of an outflow, perhaps triggered by a change in accretion state or second, fallback event.  This was suggested for the late-time X-ray-optical flares of  PS10adi, OGLE16aaa and AT2019azh, which did not have an associated radio flare. We do not feel such a model is appropriate for IGRJ12580+0134 
for two reasons.  First, in IGRJ12580+0134, the radio outflow began at early times \citep{Perlman17, Irwin15}. And second, no flare was seen in other bands for IGRJ12580+0134, as discussed above, whereas the late-time radio flare of ASSASN-15oi, was associated with an X-ray and optical flare \citep{Horesh21,Gezari17,Holoien18}. Additionally, given that the original event was associated with the accretion of a substellar object (e.g., a brown dwarf or super-jupiter \cite{Nikolajuk13}), it is unlikely that adequate mass would be left over from the addditional event.

It is quite likely that as more TDEs are discovered, an increasing diversity of ejection and flare events will be seen both in their X-ray, optical-UV and radio flux, as well as in their radio structure.  The events discussed in this paper are very likely just a preview of some of the things that will be found in the future.  Already in its first year, eROSITA has produced an impressive diversity of properties in the thirteen TDEs it discovered \citep{Sazonov21}.  This is particularly enticing with the advent of surveys such as the Rubin Observatory LSST and {\it Roman} Space Telescope coming in the near future.  Late-time radio flares, such as those found in IGRJ12580+0134 or ASASSN-15oi could be unique probes of the circumnuclear environment, particularly its density and potentially magnetic field structure (Yuan et al. 2016, Horesh et al. 2021).

\begin{acknowledgments}
The National Radio Astronomy Observatory is a facility of the National Science Foundation operated under cooperative agreement by Associated Universities, Inc. This paper was based partly on observations obtained with XMM-Newton, an ESA science mission with instruments and contributions directly funded by ESA Member States and NASA. We acknowledge the use of public data from the Swift data archive.
QY thanks Liang-Duan Liu and He Gao for helpful discussion of the modeling. \newline

The National Radio Astronomy Observatory is a function of the National Science Foundation operated under cooperative agreement by Associated Universities, Inc.
The VLA datasets were reduced by different members of the team.  EM processed the data from program 19A-425 as well as the archival data from AB0879.  TW processed the {\sl Swift}-concurrent observations of 16A-420, and YY reduced the 2015 high-resolution observations of 15A-400.

\software{CASA (v5.6.2-3; McMullin et al. 2007), SAS (Gabriel et al. 2021), Sherpa \cite{Freeman01,Doe07}}


\end{acknowledgments}

\begin{appendix}
\section{Modeling of the jet-CNM interaction}
The jet dynamics when propagating in the CNM is described by the following equations
\citep{Huang00}
\begin{eqnarray}
\frac{dR}{dt} &=& \beta c \Gamma \left(\Gamma+\sqrt{\Gamma^2-1}\right),\\
\frac{dM}{dR} &=& 2\pi R^2 \left(1-\cos\theta_j\right)n m_p,\\
\frac{d\Gamma}{dM} &=& -\frac{\Gamma^2-1}{M_{\rm ej}+\epsilon M + 2(1-\epsilon)\Gamma M},
\end{eqnarray}
where $R$ is the radius of the propagating jet, $\beta$ is the jet velocity in unit
of light speed $c$, $\Gamma=1/\sqrt{1-\beta^2}$ is the Lorentz factor of the jet,
$M$ is the swept mass of the CNM by the jet, $\theta_j$ is the jet opening angle,
$n$ is the CNM density, $m_p$ is proton mass, $M_{\rm ej}$ is the ejecta mass,
and $\epsilon$ describes the radiation efficiency which is assumed to be $\ll 1$. 
Note that we neglect the sideway expansion of the jet and thus $\theta_j$ keeps
to be constant during the jet propagation. The ejecta mass can be obtained as 
\begin{equation}
M_{\rm ej}=\frac{E_k}{(\Gamma_i-1)c^2}\frac{1-\cos\theta_j}{2},
\end{equation}
where $E_k$ is the (isotropic) kinetic energy of the ejecta and $\Gamma_i$ is the
initial Lorentz factor. The jet is assumed to propagate in a uniform CNM, and then
encounter a gas cloud with finite size and a constant density (see the parameters 
in the main text). Eqs. (A1)-(A3) are solved numerically, taking into account the 
assumed gas cloud.

After obtaining the jet dynamics, the synchrotron radiation is calculated following 
the standard afterglow emission model of gamma-ray bursts \citep{Gao13}.
For the observations at relatively low frequencies and late time which correspond to
our case, the synchrotron cooling is not important, and we have $\nu_c \gg \nu_m$,
$\nu_c \gg \nu_a$, where $\nu_c$, $\nu_m$, and $\nu_a$ are the characteristic cooling
frequency, minimum injection frequency, and self-absorption frequency, respectively.
The radiation flux density can be calculated as 
\begin{eqnarray}
F &=& F_{\rm max}\left(\frac{\nu_a}{\nu_m}\right)^{1/3}\left(\frac{\nu}{\nu_a}\right)^{2},~~~~~~~\nu<\nu_a<\nu_m, \nonumber\\
F &=& F_{\rm max}\left(\frac{\nu}{\nu_m}\right)^{1/3},~~~~~~~~~~~~~~~~~\nu_a<\nu<\nu_m, \\
F &=& F_{\rm max}\left(\frac{\nu}{\nu_m}\right)^{-(p-1)/2},~~~~~~~~~~~\nu_m<\nu, \nonumber
\end{eqnarray}
and
\begin{eqnarray}
F &=& F_{\rm max}\left(\frac{\nu_m}{\nu_a}\right)^{(p+4)/2}\left(\frac{\nu}{\nu_m}\right)^{2},~~~~~\nu<\nu_m<\nu_a, \nonumber\\
F &=& F_{\rm max}\left(\frac{\nu_a}{\nu_m}\right)^{-(p-1)/2}\left(\frac{\nu}{\nu_a}\right)^{5/2},~~\nu_m<\nu<\nu_a, \\
F &=& F_{\rm max}\left(\frac{\nu}{\nu_m}\right)^{-(p-1)/2},~~~~~~~~~~~~~~~\nu_a<\nu, \nonumber
\end{eqnarray}
where $p$ is the spectral index of accelerated electrons, $F_{\rm max}=(m_e c^2\sigma_T\Gamma 
BM)/(12\pi e d^2 m_p)$ is the maximum flux over all frequencies, with $m_e$ being the electron
mass, $\sigma_T$ being the Thomson cross section, $B$ being the magnetic field, $e$ being the 
electron charge, and $d$ being the distance of NGC 4845.

The TDE of IGR J12580+0134 is expected to be viewed from an off-axis direction of the jet
\citep{Lei16}. Some corrections of the viewing angle effect need to be applied, including
$\nu^{\rm off}=a_{\rm off}\nu^{\rm on}$, $t^{\rm off}=t^{\rm on}/a_{\rm off}$,
$F_{\nu}^{\rm off}(t)=a_{\rm off}^3F_{\nu/a_{\rm off}}^{\rm on}(a_{\rm off}\,t)$,
where $a_{\rm off}=(1-\beta)/(1-\beta\cos\theta_{\rm obs})$ \citep{Granot02}.
These corrections are important at early time before the jet decelerates to the
non-relativistic regime.

\end{appendix}


\begin{thebibliography}{}
\expandafter\ifx\csname natexlab\endcsname\relax\def\natexlab#1{#1}\fi
\providecommand{\url}[1]{\href{#1}{#1}}
\providecommand{\dodoi}[1]{doi:~\href{http://doi.org/#1}{\nolinkurl{#1}}}
\providecommand{\doeprint}[1]{\href{http://ascl.net/#1}{\nolinkurl{http://ascl.net/#1}}}
\providecommand{\doarXiv}[1]{\href{https://arxiv.org/abs/#1}{\nolinkurl{https://arxiv.org/abs/#1}}}

\end{thebibliography}


\begin{thebibliography}{1}



\bibitem[Alexander \etal (2020)]{Alexander20} Alexander, K. D., van Velzen, S., Horesh, A., Zauderer, B. A., 2020, Space Sci. Rev., 216, 81

\bibitem[Auchettl \etal (2017)]{Auchettl17} Auchettl, K., Guillochon, J., Ramirez-Ruiz, E., 2017, ApJ, 838, 149

\bibitem[Auchettl \etal (2018)]{Auchettl18} Auchettl, K., Guillochon, J., Ramirez-Ruiz, E., 2018, ApJ, 852, 37



\bibitem[Berger \etal (2012)]{Berger12} Berger, E., Zauderer, A., Pooley, G. G., 2012, ApJ, 748, 36


\bibitem[Bhatnagar \etal (2013)]{Bhatnagar13} Bhatnagar, S., Rau, U., \& Golap, K. 2013, ApJ, 770, 91

\bibitem[Blandford, Meier \& Readhead (2019)]{bland19} Blandford, R., Meier, D., Readhead, A., 2019, A\& A Rev., 57, 467 

\bibitem[Bloom \etal (2011)]{Bloom11} Bloom, J. et al. 2011, Science, 333, 203

\bibitem[Bogd\'an \& Gilfanov (2011)]{Bogdan11} Bogd\'an, A., \& Gilfanov, M., 2011, MNRAS, 418, 1901

\bibitem[Boller \etal (2003)]{Boller03} Boller, T., Voges, W., Dennefeld, M., Lehmann, I., Predehl, P., Burwitz, V., Perlman, E., Gallo, L, Papadakis, I. E., Anderson, S., 2003, A\&A, 397, 557

\bibitem[Bright \etal (2018)]{Bright18}Bright, J. S., Fender, R. P., Motta, S. E., et al., 2018, MNRAS, 475, 4011

\bibitem[Burrows \etal (2011)]{Burrows11} Burrows, D. et al. 2011, Nature, 476, 421


\bibitem[Cenko \etal (2012)]{Cenko12} Cenko, S. et al. 2012, ApJ, 753, 77





\bibitem[Condon \etal (1998)]{Condon98} Condon, J. J., Cotton, W. D., Greisen, E. W., Yin, Q. F., Perley, R. A., Taylor, G. B., \& Broderick, J. J. 1998, AJ, 115, 1693 


\bibitem[Dickey \& Lockman (1990)]{Dickey90} Dickey, J. M., \& Lockman, F. J., 1990, ARAA, 28, 215

\bibitem[Doe, Nguyen \& Stawarz (2007)]{Doe07} Doe, S., Nguyen, D., Stawarz, C., et al., 2007, in ADASS XVI, ASP Conference Series, v. 376, ed. R. A. Shaw, F. Hill \& D. J. Bell (ASP:  San Francisco), p. 543

\bibitem[Freeman, Doe \& Siemiginowka (2001)]{Freeman01} Freeman, P., Doe, S., Siemiginowska, A., 2007, Proc. SPIE, 4477. 76

\bibitem[Gabriel (2021)]{Gabriel21} Gabriel, C., "Users Guide to the XMM-Newton Science Analysis System", Issue 16.0, 2021 (ESA: XMM-Newton SOC).

\bibitem[Gao \etal (2013)]{Gao13} Gao, H., Lei, W. H., Zou, Y. C., Wu, X. F., \& Zhang, B. 2013, NewAR, 57, 141



\bibitem[Gezari (2021)]{Gezari21} Gezari, S., 2021, A\&A Rev., in press (arXiv: 2104.14580)

\bibitem[Gezari \etal (2017)]{Gezari17} Gezari, S., et al., 2017, ApJ, 851, L47
  
\bibitem[Granot \etal (2002)]{Granot02} Granot, J., Panaitescu, A., Kumar, P., \& Woosley, S.~E. 2002, \apjl,  570, L61


\bibitem[Hervet \etal (2017)]{hervet17} Hervet, O., Meilani, Z., Zech, A., Boisson, C., Cayatte, V., Sauty, C., Sol, H., 2017, A\& A, 606, 103

\bibitem[Ho \etal (1995)]{Ho95} Ho, L. C., Filippenko, A. V., Sargent, W. L. W., 1995, ApJS, 98, 477
 



\bibitem[Holoien \etal (2016)]{Holoien16} Holoien, T. W. S., Kochanek, C. S., Prieto, J. L., et al., 2016, MNRAS, 455, 2918

\bibitem[Holoien \etal (2018)]{Holoien18} Holoien, T. W. S., Brown, J. S., Auchettl, L., et al., 2018, MNRAS, 480, 5689

\bibitem[Horesh, Cenko \& Arcavi (2021)]{Horesh21} Horesh, A., Cenko, S. B., Arcavi, I., 2021, Nature Astronomy, 5, 491

\bibitem[Huang \etal (2000)]{Huang00} Huang, Y.~F., Gou, L.~J., Dai, Z.~G., \& Lu, T. 2000, \apj, 543, 90

\bibitem[Huang \etal (2021)]{Huang21} Huang, S., Hu, S., Yin, H., Chen, X., Alexeeva, S., Gao, D., Jiang, Y. 2021, ApJ submitted, arXiv:2106.14368

\bibitem[Irwin \etal (2012)]{Irwin12} Irwin, J., Beck, R., Benjamin, R.~A., et al.\ 2012, \aj, 144, 43 

\bibitem[Irwin \etal (2013)]{Irwin13} Irwin, J., Krause, M., English, J., et al.\ 2013, \aj, 146, 164 

\bibitem[Irwin \etal (2015)]{Irwin15} Irwin, J. et al. 2015, ApJ, 809, 172



\bibitem[Jia \etal (2012)]{Jia12} Jia, J., Ptak, A., Heckman, T. M., Braito, V., \& Reeves, J., 2012, ApJ, 759, 41

\bibitem[Jiang \etal (2019)]{Jiang19} Jiang, N., Wang, T., Yan, L.,Xiao, T., Yang, C., Dou, L., Wang, H., Cutri, R., Mainzer, A., 2017, ApJ, 850, 63

\bibitem[Kajava \etal (2020)]{Kajava20} Kajava, J. J. E., Guistini, M., Saxton ,R. D., Miniutti, G.,  2020, A\& A, 639, A100

\bibitem[Kara \etal (2016)]{Kara16} Kara, E., Miller, J. M., Reynolds, C, Dai, L., 2016, Nature, 535, 388 


\bibitem[Lacy \etal (2019)]{Lacy19} Lacy, M., Chandler, C., Kimball, A., Myers, S., Nyland, K., \& Witz, S. 2019, in Astronomical Data Analysis Software and Systems XXVII, ed. P. J. Teuben, M. W. Pound, B. A. Thomas, \& E. M. Warner, ASP, Conf. Series, 523, 217

\bibitem[Larionov \etal (2020)]{Larionov20} Larionov, V. M., Jorstad, S. G., Marscher, A. P., et al., 2020, MNRAS, 492, 3829

\bibitem[Lei \etal (2016)]{Lei16} Lei, W., Yuan, Q., Zhang, B., Wang, D. 2016, ApJ, 816, 20

\bibitem[Levan \etal(2011)]{Levan11} Levan, A. J., Tanvir, N. R., Cenko, S. B., et al., 2011, Science, 333, 199

\bibitem[Liu \etal (2020)]{Liu20} Liu, L. D., Gao, H., Zhang, B. 2020, ApJ, 890, 102

\bibitem[Liu \etal (2019)]{Liu19} Liu, X.-L., Dou, L.-M., Shen, R.-F., Chen, J.-H., 2019, arXiv:1912.06081

\bibitem[Lodato \etal (2009)]{Lodato09} Lodato, G., King, A. R., Pringle, J. E., 2009, MNRAS, 392, 332

\bibitem[Lodato \& Rossi (2011)]{lodato11} Lodato, G., Rossi, E., 2011, MNRAS, 410, 359


\bibitem[Magorrian \& Tremaine (1999)]{Magorrian99} Magorrian, J. \& Tremaine, S. 1999, MNRAS, 309, 447

\bibitem[Mattila \etal (2018)] {Mattila18} Mattila, S., et al., 2018, Science, 361, 482

\bibitem[McMullin \etal (2007)]{McMullin07} McMullin, J. P., Waters, B., Schiebel, D., Young, W., Golap, K., 2007, in ADASS XVI, ASP Conf. Series, 376, ed. Shaw, R. A., Hill, F., and Bell, D. J., (ASP:  San Francisco), 127



\bibitem[Neustadt \etal (2020)]{Neustadt20} Neustadt, J. M. M., Holoien, T. W. S., Kochanek, C. S., et al. 2020, MNRAS, 484, 2538

\bibitem[Nikolajuk \& Walter (2013)]{Nikolajuk13} Nikolajuk, M. \& Walter, R. 2013, A\&A, 552, A75

\bibitem[Onori, Cannizzaro, Jonker \etal (2019)]{Onori19} Onori, F., Cannizzaro, G., Jonker, P. G., et al., 2019, MNRAS, 489, 1463



\bibitem[Paliya \etal (2019)]{Paliya19} Paliya, V. S., Koss, M., Trakhtenbrot ,B. 2019, ApJ, 881, 154

\bibitem[Pasham \& van Velzen (2018)]{Pasham18} Pasham, D. R., van Velzen, S., 2018, ApJ, 856, 1

\bibitem[Perley \& Butler (2013)]{Perley13} Perley, R. A., Butler, B. J., 2013, ApJS, 204, 19

\bibitem[Perlman (2013)]{perl13} Perlman, E. S., 2013, in Planets, Stars and Stellar Systems, Vol. 6 (Springer: Dordrecht), p.305

\bibitem[Perlman \etal (1999)]{Perlman99} Perlman, E. S., Madejski, G., Stocke, J. T., Rector, T. A., 1999, ApJ, 523, L11

\bibitem[Perlman \etal (2005)]{Perlman05} Perlman, E. S., Madejski, G., Georganopoulos, M., et al., 2005, ApJ, 625, 727

\bibitem[Perlman \etal (2017)]{Perlman17} Perlman, E. S., Meyer, E. T., Wang, Q. D., et al., 2017, ApJ, 842, 126

\bibitem[Phinney (1989)]{Phinney89} Phinney, E. 1989, IAU Symposium, 136, 543

\bibitem[Ptak et al. (1999)]{Ptak99} Ptak, A., Serlemitsos, P., Yaqoob, T., \& Mushotzky, R. 1999, ApJS, 120, 179

\bibitem[Rau \& Cornwell (2011)]{Rau11} Rau, U., \& Cornwell, T. J. 2011, \aap, 532, A71


\bibitem[Ricci \etal (2017)]{Ricci17} Ricci, C., Trakhtenbrot, B., Koss, M. J., et al., ApJS, 233, 17



\bibitem[Saxton \etal (2021)]{Saxton21} Saxton, R., Komossa, S., Auchettl, K., Jonker, P. G., 2021, Space Sci. Rev. 217, 18

\bibitem[Sazonov, Gilfanov, Medvedev \etal (2021)]{Sazonov21}Sazonov, S., Gilfanov, M., Medvedev, P., Yao, Y., Khorunzhev, G., Semena, A., Sunyaev, R., Burenin, R., Lyapin, A.,, Mescheryakov, A., Uskov, G., Zaznobin, I., Postnov, K.A., Dodin, A.V., Belinski, A.A., Cherepashchuk, A.M., Eselevich, M., Dodonov S.N., Grokhovskaya, A.A., Kotov, S.S., Bikmaev I.F., Zhuchkov, R.Y, Gumerov R.I., van Velzen S., Kulkarni S., MNRAS, in press, arXiv: 2108.02449
4

\bibitem[Shen (2021)]{Shen21} Shen, Y., 2021, ApJ, in press, arXiv:2108.05381]




\bibitem[Stark \etal (1992)]{Stark92} Stark, A. A., Gammie, C. F., Wilson, R. W., et al., 1992, ApJS, 79, 77

\bibitem[Stolc \& Karas (2019)]{Stolc19} Stolc, M., Karas, V., 2019, Astron. Nachr., 340, 570

\bibitem[Subroweit, Garcia-Marin, Eckart \etal (2017)]{Subroweit17} Subroweit, M., Garcia-Marin, M., Eckart, A., Borkar, A., Valencia, M., Witzel, G., Shahzamanian, B., Straubmeier, C., 2017, A\& A, 601, A80.

\bibitem[Trakhtenbrot \etal (2019a)]{Trakhtenbrot19a} Trakhtenbrot, B., Arcavi, I., MacLeod, C. L., 2019, ApJ, 883, 94
 
\bibitem[Trakhtenbrot \etal (2019b)]{Trakhtenbrot19b} Trakhtenbrot, B., Arcavi, I., Ricci, C., 2019,  Nat. Astron., 3, 242 



\bibitem[van Velzen \etal (2019)]{vanVelzen19} van Velzen S., Stone N. C., Metzger B. D., Gezari S., Brown T. M., Fruchter A. S. 2019, ApJ, 878, 82

\bibitem[Urry \& Shafer (1984)]{Urry84} Urry, C. M., \& Shafer, R. A., 1984, ApJ, 280, 569

\bibitem[Van Velzen \& Farrar (2014)]{vanVelzen14} Van Velzen, S., Farrar, G., 2014, ApJ, 792, 53

\bibitem[Walter \etal (2011)]{Walter11} Walter, R. et al. 2011, ATel, 3108

\bibitem[Wang \& Merritt (2004)]{Wang04} Wang, J. \& Merritt, D. 2004, ApJ, 600, 149


\bibitem[Weaver et al. (2018)]{Weaver18} Weaver, K., Valencic, L., Perry, B., Arida, M., Kuntz, K., Snowden, S., Harrus, I., Immler, S., Shafer, R., Smith, R., Still, M., The {\sl XMM-Newton} ABC Guide:  An Introduction to {\sl XMM-Newton} Data Analysis, v 6.0, 2018, NASA/GSFC {\sl XMM-Newton} Guest Observer Facility, online at https://heasarc.gsfc.nasa.gov/docs/xmm/abc/

\bibitem[Wiegert \etal (2015)]{Wiegert15} Wiegert, T., Irwin, J., Miskolczi, A., et al.\ 2015, \aj, 150, 81 




\bibitem[Yalinewich \etal (2019)]{Yalinewich19} Yalinewich,A., Steinberg,E., Piran,T, Krolik,J.H., 2019, MNRAS, 487, 4083

\bibitem[Yuan \etal (2016)]{Yuan2016} Yuan, Q., Wang, Q.~D., Lei, W.-H., Gao, H., \& Zhang, B. 2016, \mnras, 461, 3375

\bibitem[Zabludoff \etal (2021)]{Zabludoff21}Zabludoff, A., Arcavi, I., La Massa, S., et al., 2021, Space Sci. Rev. 217, 54

\bibitem[Zauderer \etal (2011)]{Zauderer11} Zauderer, B. A., et al. 2011, Nature, 476, 425

\bibitem[Zauderer \etal (2013)]{Zauderer13} Zauderer, B. A., Berger, E., Margutti, R., et al., 2013, ApJ, 767, 152

\end{thebibliography}
\end{document}